\documentclass[dvips]{elsart}
\usepackage[intlimits,sumlimits]{amsmath}
\usepackage{amssymb,amsfonts,amsgen,amsbsy,amscd,amsopn,amstext,amsxtra}
\usepackage[final]{graphicx}
\usepackage[applemac]{inputenc} 
\usepackage[comma,sort&compress,square]{natbib}

\allowdisplaybreaks

\newcommand{\english}[1]{#1}
\newcommand{\deutsch}[1]{}

\newcommand{\deutschnospace}[1]{}
\newcommand{\nopaper}[1]{}
\newcommand{\onlypaper}[1]{#1}

\newcommand{\real}{\mathbb R}
\newcommand{\ableit}[2]{\frac{\partial #1}{\partial #2}}

\newcommand{\sig}[1]{\vec\sigma_{#1}}

\newcommand{\Fourier}{\mathcal F} 
\renewcommand{\(} {\left (}
\renewcommand{\)} {\right )}
\newcommand{\fig}[1]{\autoref{#1}} 
\newcommand{\zit}[1]{(\ref{#1})}
\newcommand{\equ}[1]{%
\deutsch{Gleichung}%
\english{equation}%
~\zit{#1}}
\newcommand{\autoref}[1]{section \ref{#1}} 

\newcommand{\nameequ}[2]{#1~(\ref{#2})}
\newcommand{\oveq}[1]{\overset{ #1}=}
\newcommand{\ovapprox}[1]{\overset{ #1}\approx}
\newcommand{\oveqzit}[1]{\oveq{\zit{#1}}}
\newcommand{\ovun}[3]{\underset{#2}{\overset{#1}#3}}
\newcommand{\ovuneq}[2]{\ovun{#1}{#2}{=}}
\newcommand{\ovuneqzit}[2]{\ovuneq{\zit{#1}}{\zit{#2}}}
\newcommand{\ovdefeq}{\oveq{\text{Def.}}}
\newcommand{\cont}{{\otimes}}
\newcommand{\win}{{\mathcal W}}
\begin{document}
\begin{frontmatter}
\title{Radiation spectra of laser-driven quantum relativistic electrons}
\author{Guido R. Mocken}\ead{mocken@physik.uni-freiburg.de}
\and
\author{Christoph H. Keitel}\ead{keitel@mpi-hd.mpg.de}
\address{Max-Planck-Institut f\"ur Kernphysik, Saupfercheckweg 1, D-69117 Heidelberg, Germany}
\address{Theoretische Quantendynamik, Physikalisches Institut, Universit\"at Freiburg,\\
Hermann-Herder-Stra{\ss}e 3, D-79104 Freiburg, Germany}

\begin{abstract}
A procedure to calculate the radiation spectrum emitted by an arbitrarily prepared Dirac wave packet is developed. It is based on the Dirac charge current and classical electrodynamic theory. Apart from giving absolute intensity values, it is exact in terms of relativistic retardation effects and angular dependence.
We employ a laser driven free electron to demonstrate the advantages of our method as compared to traditional ones that merely rely on the Fourier transform of the dipole operator's expectation value. Classical reference calculations confirm the results obtained for the low-frequency part of the spectrum, especially in terms of the observed red-shifts, which clearly deviate from non-relativistic calculations. In the high-frequency part of the spectrum, we note appreciable deviations to the purely classical calculations which may be linked to quantum averaging effects.
\end{abstract}
\end{frontmatter}
%
\section{Introduction}
\label{intro}
\english{The appearance of harmonics in the light scattered at free electrons is known both in theory  \cite{Vachaspati:1962,Brown:1964,Sarachik:1970,Salamin:1996} as well as in experiments  \cite{Englert:1983,Chen:1998,Chen:2000} for a rather long time now. Atomic systems  \cite{Bandarage:1992,Keitel:1994,SzymanowskiKeitel:1999,KeitelSzymanowski:1998,Wagner:1999,Prager:2001} in highly intense laser fields have turned out to produce especially high orders, the so-called High Harmonics \cite{Joachain:2000,Brabec:2000,Protopapas_review:1996}. In this case, the experimentally observed spectra  \cite{McPherson:1987,Ferray:1988,Preston:1996} are usually explained in terms of the so-called recollision model  \cite{Lewenstein:1994}. Therefore, in the context of quantum mechanical simulations such as those shown in \cite{Mocken:2003,Mocken:2003b,Mocken:2003c}, apart from the obvious observables such as the expectation value of the position or momentum operator and the spatial probability density distribution, there is a lot of interest in calculating the radiation spectrum as well.}%

\english{The numerical methods that were described in our previous work \cite{Mocken:2003b} provide the Dirac wave function as a function of  time and position, where both variables are discretized and limited to certain regions. There exists a well-established procedure \cite{Burnett:1992} that derives the emission spectrum from such a given data set (e.g. \cite{KeitelHu:2001,Prager:2001}). However, this one was developed for the non-relativistic Schr\"odinger theory, and contains further approximations. 
The question arises whether, in the highly relativistic regime, where non-relativistic quantum theory breaks down, it is still advisable to employ a partly non-relativistic method for the calculation of the quantum emission spectrum.}

\english{In this paper, we present a procedure that takes into account all relativistic spectral effects. It is based on the Dirac charge current and the classical theory of electromagnetism. Apart from giving absolute intensity values, it is exact in terms of relativistic retardation effects and angular dependence.
To take advantage of the new scheme, one needs to obtain the time-resolved Dirac wave function of some physically interesting scenario in the first place. 
We made use of our advanced two-dimensional Dirac split-operator code \cite{Mocken:2003b} and considered the motion of a free electron in a laser field \cite{Roman:2000} to put the new scheme to the test. Classical calculations, which should not deviate much from the quantum calculations for this particular case as far as the lower part of the spectrum is concerned, serve as a reference and prove the correctness of the new scheme. However, significant deviations from the established method \cite{Burnett:1992} are revealed. In addition to that, the higher order harmonics are averaged out. The delocalized quantum charge distribution is likely to be responsible for this effect.}

This remainder of this paper is organized as follows:
\english{Section \ref{sec:class_rad} briefly reviews the formulas required to evaluate the radiation spectrum of a classical point charge and their non-relativistic limit. Section \ref{sec:qm_rad} briefly discusses the established standard methods, then turns to  the derivation of our new method. In the end, the non-relativistic limit of the latter is evaluated to show that it matches the former.
This section also discusses the solution of some numerical problems that are involved.
Section \ref{chap:results_spectrum} discusses the results obtained with our method and relates them to those obtained with the standard methods and classical formulas. We draw our conclusions in section \ref{sec:conclusions} and provide the classical equations of motion in appendix \ref{chap:rk4} and the Dirac equation in appendix \ref{sec:Dirac_Hestenes}, as well as some remarks and references concerning the way that we solved both problems numerically.}

\section{Classical radiation spectrum}
\label{sec:class_rad}
\english{The physical quantity of interest here is the differential amount of energy ($d^2W$) emitted into a certain interval of  frequency ($d\omega$) and solid angle ($d\Omega$), i.e. the quantity $\frac{d^2W}{d\omega d\Omega}$.}

\subsection{Spectrum of a charged classical point particle in relativistic motion}
\label{sec:cm_spectrum}
\english{From \cite{book:Jackson_dt} we obtain a formula for the far-field radiation of a classical charged point particle which is ready for application:}
\begin{eqnarray}
			\frac{d^2W}{d\omega d\Omega}\left(\vec x, \omega \right) &=& \frac{q^2}{4\pi^2c}\left|\int_{-\infty}^{+\infty} \mkern-10mu dt \frac{\vec n \times \left(\left(\vec n - \vec \beta\right)\times\dot{\vec\beta}\right)}{\left(1-\vec n \cdot \vec \beta\right)^2}  e^{i\omega\left(t +\frac{R}{c} \right)}\right|^2,\nonumber\\
\text{\english{where}\deutsch{mit}~}
			R(t) &=& \left| \vec x - \vec y(t)\right|, \quad
			\vec n(t) =  \frac{\vec x - \vec y(t)}{R(t)}, \quad
			\vec \beta(t) = \frac{\dot {\vec y}(t)}{c}=\frac{1}{c}\frac{dy}{dt}(t).
\label{eq:d2Wclassical}
\end{eqnarray}
\english{Here, $\vec y(t)$ represents the particle's trajectory and $\vec x$ the observer's position. The particle's charge is labelled $q$, and $c$ is the speed of light.}
\english{The integral $\int dt (\ldots)$ is replaced by a sum $\sum \Delta t (\ldots)$ which is carried out for all values $\omega$ that are of interest by using the discrete trajectory data for $\vec y, \dot{\vec y}, \ddot{\vec y}$, that is available from the numerical integration of the classical equation of motion  \zit{eq:DGLsytem}.}
\subsection{Non-relativistic limit}
\label{sec:cm_spectrum_nonrel}
\english{If one omits relativistic corrections ($|\vec \beta|\ll 1$) and accounts (with respect to the retardation $R$) only for particle displacements in the direction towards the observer that are very small ($\hat x \cdot \vec y\approx 0$) as compared to the distance of the fixed observer $|\vec x|$, then $R$ is constant, }%
\begin{equation}
R = \left| \vec x - \vec y\right| \oveq{|\vec y| \ll |\vec x|} |\vec x| - \hat x \cdot \vec y + O\left(\frac{|\vec y|^2}{|\vec x|}\right)\approx const.  - \hat x \cdot \vec y \ovapprox{\hat x \cdot \vec y\approx 0} const.\label{eq:RetApprox}
\end{equation}%
\english{and instead of the integral in \equ{eq:d2Wclassical}, the result is just the Fourier transform of the acceleration component $\vec a_{\perp}$ that is perpendicular to the momentary direction of observation $\vec n$:}
\begin{eqnarray}
	\frac{d^2W}{d\omega d\Omega}\left(\vec x, \omega \right) &\ovuneq{\nopaper{\zit{eq:RetApprox}}}{\nopaper{|\vec \beta|\ll 1}}& \frac{q^2}{4\pi^2c}\left|\int_{-\infty}^{+\infty} \mkern-10mu dt  \;\vec n \times \left(\vec n \times{\dot{\vec\beta}}\right)  e^{i\omega t  }\right|^2 = \frac{q^2}{2\pi c}\left|\frac{-1}{\sqrt{2\pi}}\int_{-\infty}^{+\infty} \mkern-10mu dt  \; \dot{\vec\beta}_{\perp(t)}  e^{i\omega t  }\right|^2\nonumber\\
&=& \frac{q^2}{2\pi c}\left| -\dot{\vec\beta}_{\perp(-\omega)}  \right|^2=\frac{q^2}{2\pi c}\left| \dot{\vec\beta}_{\perp(\omega)}  \right|^2.
\end{eqnarray}
\english{Using $\dot{\vec\beta}_{\perp} \ovdefeq \frac{\vec a_{\perp}}{c}$ one finally obtains:}
{\begin{equation}
\frac{d^2W}{d\omega d\Omega}\left(\vec x, \omega \right) =  \frac{q^2}{2\pi c^3}\left|  \vec a_{\perp(\omega)}  \right|^2  \label{eq:d2Wclassical_nonrel}.
\end{equation}}
\section{Quantum mechanical treatment}
\label{sec:qm_rad}
\english{The usual way (e.g. \cite{KeitelHu:2001,Prager:2001}) to calculate the emission spectrum is the following: Starting from the classical non-relativistic formula \zit{eq:d2Wclassical_nonrel} }%
\english{the spectrum for a quantum mechanical system is evaluated using the corresponding quantity that is based on the acceleration expectation value $\langle \vec a \rangle$:}%
\begin{equation}
\frac{d^2W}{d\omega d\Omega}\left(\vec x, \omega \right)  =  \frac{q^2}{2\pi c^3}\left|  \langle \vec a \rangle_{\perp(\omega)}  \right|^2
\end{equation}
\english{For a more precise derivation and in addition to that a detailed discussion of the role of the ``coherent'' part (see above) and the  ``incoherent'' part (not discussed here) of the spectrum, see \cite{Eberly:1989,Eberly:1992}. The next step \cite{Burnett:1992} is the application of the Ehrenfest theorem, i.e. the replacement}%
\begin{equation}
m \langle \vec a \rangle  \ovuneq{\text{Ehrenfest}}{\text{Newton}} \langle \vec F\rangle \ovuneq{\text{Lorentz-}}{\text{\deutsch{kraft}\english{force}}} \langle q \vec E + \frac{q}{c}\vec v\times \vec B\rangle, \label{eq:Newton_Ehrenfest}
\end{equation}
\english{which is easy to evaluate if the magnetic field  $\vec B$ is neglected (the so-called dipole approximation). However, it is the magnetic field effects and the forward drift caused by them which have to be taken care of in this work with an emphasis on relativistic velocities. This precludes \equ{eq:d2Wclassical_nonrel} from being the basis of any highly-relativistic theory.}
\english{In a weakly-relativistic work such as  \cite{Hu:2001}, the starting point remains  \equ{eq:d2Wclassical_nonrel}, however, the relativistically correct \equ{eq:DGLsystem1} is employed instead of \equ{eq:Newton_Ehrenfest}. Utilising several essential approximations, the authors obtain an expression that can be evaluated numerically. Those approximations are no longer valid in the more strongly relativistic cases considered here, where even for classical calculations, the substitution of \equ{eq:d2Wclassical} by \zit{eq:d2Wclassical_nonrel} can already lead to noticeable frequency shifts in the spectrum.}%

\english{Unfortunately, in a quantum mechanical context, \equ{eq:d2Wclassical_nonrel} is not easily replaced by the relativistically correct variant  \zit{eq:d2Wclassical} and a calculation of the required expectation values in the framework of the Dirac theory. On the one hand there are fundamental reasons, such as the difficulty to define a reasonable velocity operator in the Dirac theory\footnote{See the discussion of Zitterbewegung and the Foldy-Wouthuysen transform in \cite{book:Strange}.}, on the other hand one needs to take into account certain practical considerations\footnote{Operator functions that contain quotients and square roots, and in which mixed  position and momentum operator dependent terms occur (symmetrized as necessary), can be evaluated only with disproportionately large numerical effort.}. Therefore we have dropped all corresponding considerations in favour of the following far more  promising idea.}%

\english{We start by calculating the emission spectrum of a spatially distributed classical charge current, instead of the one of a point charge. The following derivation differs slightly from those found in the literature  \cite{book:Jackson_dt} in that the one and only approximation made is the far-field expression for the electric field of a charge current. Otherwise it is exact. The transition to quantum mechanics is carried out at the very end of the calculation by substituting a suitable expression for the charge current.}%
\subsection{Fully relativistic spectrum of a spatially distributed charge current}
\label{sec:qm_spectrum}
\begin{figure}[p]
\begin{center}
	\includegraphics[width=\textwidth]{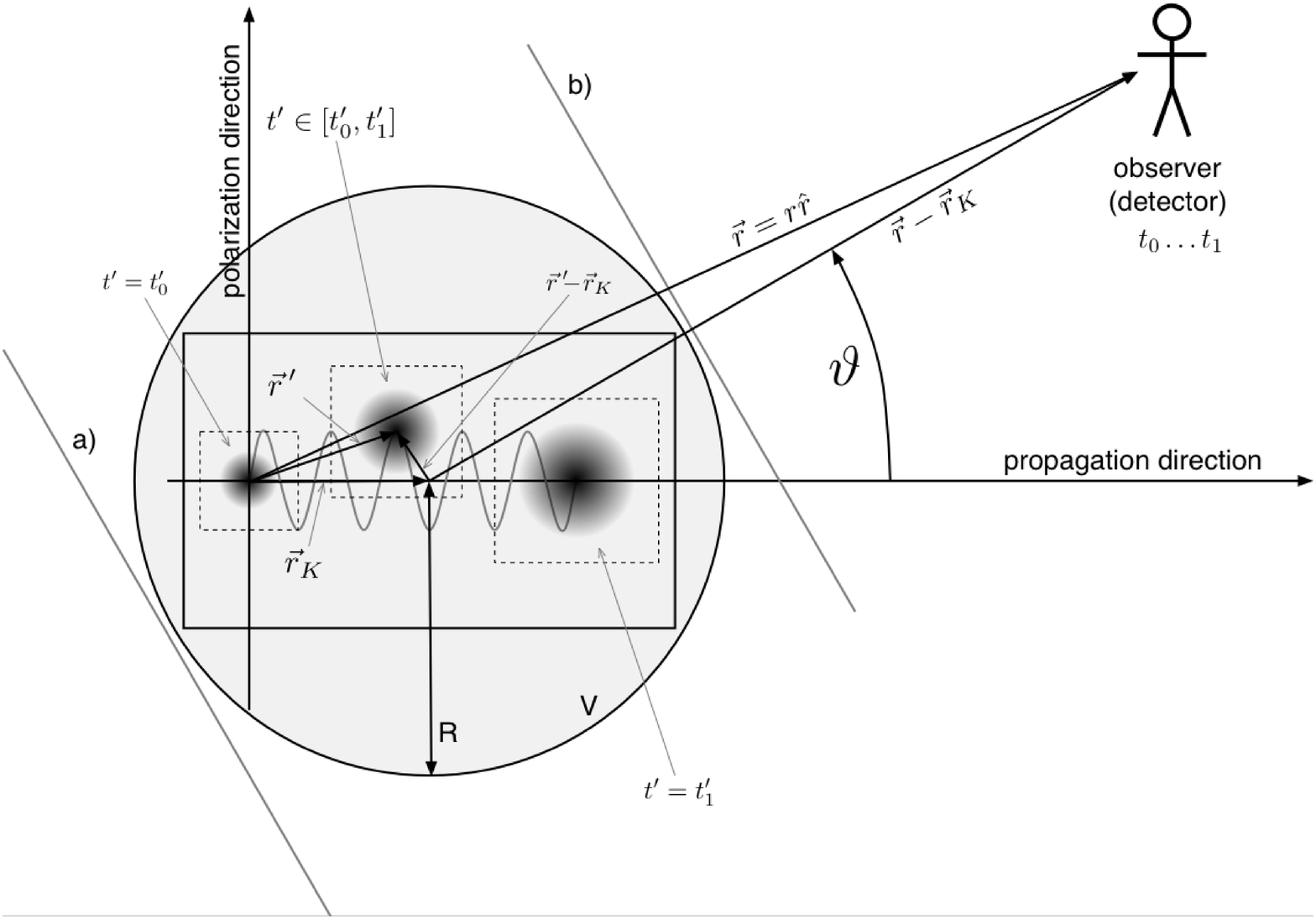}
\caption[Skizze]{%
\english{Sketch of the radiation scenario: The three black-shaded areas that carry time ($t'$)  labels represent the electron at the beginning ($t'=t_0'$) and the end ($t'=t_1'$) of the simulation, as well as at an arbitrary point in time ($t'\in [t_0',t_1']$) in between, each enclosed inside the dashed boundaries of the co-moving and -growing numerical grid.
The oscillatory curve is the trajectory of the particle's centre. The observer (or detector),  which is activated during the time interval $t_0\ldots t_1$, is located  under an angle  $\vartheta$ with respect to the laser propagation direction at a distance $|\vec r-\vec r_k|$  as measured from the middle of the volume $V$ (large grey-shaded circle), which is a sphere of radius $R$ and origin $\vec r_K$. At the same time, the radius $R$ represents half the diagonal line of the solid rectangle, which is the enclosing envelope of all occurring numerical grids. The two parallel lines perpendicular to the line of observation labelled ``a)'' und ``b)'' mark two specific distances to the observer mentioned in the text.}%
}
\label{fig:Skizze_Spektrum}
\end{center}
\end{figure}

\english{
If we call the magnitude of the Poynting-vector (energy per area and unit time) $|\vec S|$, its direction $\vec n$, and the electric and magnetic field $\vec E$ and $\vec B$, then we obtain}
\begin{eqnarray}
\vec S &=& \frac{c}{4\pi} (\vec E \times \vec B)\oveq{\vec B=\vec n \times \vec E} \frac{c}{4\pi} (\vec E \times ( \vec n \times \vec E))
	\oveq{\vec n \cdot \vec E=0} \frac{c}{4\pi} \vec E^2 \vec n \nonumber\\* & \Rightarrow& |\vec S| = \frac{c}{4\pi} \vec E^2\label{eq:Smag}
\end{eqnarray}
\english{We assume the radiation detector to be located at position  $\vec r=r \hat r$ and to be activated during the time interval $[t_0, t_1]$. Let  $\win(t)$ be a function\footnote{Later on, we will restrict ourselves to the simple case $h(t)=1$.} that describes the detection interval: }
\begin{equation}
\win(t) = \begin{cases}
     h(t)>0 &\forall t \in[t_0, t_1], \\
     0     & \text{\english{otherwise}\deutsch{sonst}}.
\end{cases} \label{eq:window}
\end{equation}
\english{If we designate  energy by $W$, power by $P$ and angular frequency by $\omega$, then the energy per unit solid angle  $\Omega$ collected by the detector at position $\vec r$ in the time interval of interest $[t_0, t_1]$ is }
\begin{eqnarray}
\frac{dW}{d\Omega}(\vec r\,) &\oveqzit{eq:window}& \int_{-\infty}^{+\infty} \mkern-10mu  dt\;\; \win(t) \frac{dP}{d\Omega}(t)\\
 	&\ovuneq{dP=|\vec S| r^2 d\Omega}{\zit{eq:Smag}}& \int_{-\infty}^{+\infty}  \mkern-10mu dt\; \;\win(t) \frac{c}{4\pi} r^2 (\vec E (t, \vec r\,))^2 \label{eq:d1Wa}.
\end{eqnarray}
\english{Here, we insert the Fourier transform of the electric field and its inverse:}
\begin{eqnarray}
\vec E (\omega, \vec r\,) &\nopaper{\oveqzit{eq:fourier_EM}}\onlypaper{=}& \frac{1}{\sqrt{2\pi}}   \int_{-\infty}^{+\infty} \mkern-10mu dt  \, \vec E(t, \vec r\,) e^{-i\omega t}\ovdefeq\Fourier \(\vec E_{(t, \vec r\,)}\)_{(\omega, \vec r\,)} \label{eq:fourier}\\
\vec E (t, \vec r\,)&\nopaper{\oveqzit{eq:fourierInv_EM}}\onlypaper{=}& \frac{1}{\sqrt{2\pi}}   \int_{-\infty}^{+\infty} \mkern-10mu d\omega  \, \vec E(\omega, \vec r\,) e^{i\omega t}\ovdefeq\Fourier^{-1}\( \vec E_{(\omega, \vec r\,)} \)_{(t, \vec r\,)}. \label{eq:fourierInv}
\end{eqnarray}%
\english{Since $\sqrt{\win} \vec E$ is real, we have $\Fourier \(\sqrt{\win} \vec E\)_{(\omega) }=\Fourier \(\sqrt{\win} \vec E\)^\dagger_{(-\omega)}$ and therefore one finally arrives at}
\begin{equation}
\frac{dW}{d\Omega}(\vec r\,)=   \nopaper{\frac{c}{4\pi} r^2    \int_{-\infty}^{+\infty} \mkern-10mu d\omega  \,    \left\{\Fourier \(\sqrt{\win} \vec E\)_{(\omega)}\right\}^\dagger   \cdot\Fourier \(\sqrt{\win} \vec E\)_{(\omega)}   \nonumber\\
 &\oveqzit{eq:kplx_btrg}&}  \frac{c}{2\pi} r^2    \int_{0}^{+\infty} \mkern-10mu d\omega'  \,   \left | \Fourier \(\sqrt{\win} \vec E\)_{(\omega')} \right |^2. \label{eq:d1W} 
\end{equation}
\english{The integrand of \zit{eq:d1W} yields}
\begin{equation}\label{eq:d2W}
\frac{d^2W}{d\omega d\Omega}= \frac{c}{2\pi} \left | r \Fourier \(\sqrt{\win} \vec E\)_{(\omega)} \right |^2.
\end{equation}
\english{The electric field itself is given by the time derivative of a certain vector potential $\vec A_f$ as obtained in the usual far field approximation for a classical charge current}
$\vec \jmath\, (t', \vec r\,')$ \english{that is essentially localized at the origin of the coordinate system:}
\begin{eqnarray}
\vec E(t, \vec r\,) &\ovuneq{\text{\english{far}\deutsch{Fern-}}}{\text{\english{field}}}& - \frac{1}{c}\ableit{}{t} \vec A_f(t, \vec r\,), \quad\text{\english{where}}\label{eq:eFeld}\\
	\vec A_f (t, \vec r\,) &\ovdefeq& \frac{1}{cr} \int_{\mathbb R^3} \mkern-3mu d^3 r'  \; \hat r \times \{\vec \jmath\,(ct-r+\hat r \cdot \vec r\,', \vec r\,') \times \hat r \}	+ O\(\frac{1}{r^2}\). \label{eq:aFeld}
\end{eqnarray}
\english{Here, $\hat r$ is a constant unit vector that points towards the observer who is situated at a fixed distance $r$, i.e. at position $\vec r =r \hat r$.}
\english{If we insert \zit{eq:aFeld} into \zit{eq:eFeld}, convert the result to the frequency domain using \zit{eq:fourier} and, while keeping in mind \zit{eq:window}, insert everything into \zit{eq:d2W}, we finally obtain, neglecting contributions that vanish as $O\(\frac{1}{r}\)$ for large $r$}
\begin{eqnarray}
\frac{d^2W}{d\omega d\Omega}\mkern-0mu &\oveq{r_0=ct}& \frac{1}{4\pi^2c^3} \Biggl | \, \int_{c t_0}^{c t_1}\!\! \!dr_0  \! \int_{\mathbb R^3} \mkern-8mu d^3 r'  \! \sqrt{h_{(\frac{r_0}{c})}} \! \(\hat r \times \left\{ \ableit{}{r_0}\vec \jmath\,(r_0-r+\hat r \cdot \vec r\,', \vec r\,') \times \hat r \!\right\} \) \cont \nonumber\\ & &\qquad\qquad\cont \qquad e^{-\frac{i\omega r_0}{c}}\Biggr |^2\label{eq:vorlRes}.
\end{eqnarray}
\english{In the above equation and some more to follow the symbol ``$\cont$'' is just a multiplicative line continuation marker, which we have to distinguish from the cross product ``$\times$''.}
\english{Note that so far, the calculation is an entirely classical one. We are aiming at replacing the classical current by the quantum mechanical one. Since our numerical Dirac code delivers $\psi$ and hence $\vec \jmath$ at fixed, equally spaced times, whereas  \equ{eq:vorlRes} requires very different times depending on position $\vec r\,'$, we have to rewrite the above as follows: We substitute}
\begin{eqnarray}
r_0'&=&r_0 - r + \hat r \cdot \vec r\,' \label{eq:subs1}\\
\frac{d^2W}{d\omega d\Omega}& \ovuneqzit{eq:subs1}{eq:vorlRes}&  \frac{1}{4\pi^2c^3} \Biggl |\,    \int_{\mathbb R^3} \mkern-8mu d^3 r' \mkern-20mu \int_{c t_0 - r + \hat r \cdot \vec r\,'}^{c t_1 - r + \hat r \cdot \vec r\,'} \mkern-30mu dr_0'  \! \sqrt{h_{\(\frac{r_0+ r - \hat r \cdot \vec r\,'}{c}\)}} \! \(\hat r \times \left\{ \ableit{}{r_0'}\vec \jmath\,(r_0', \vec r\,') \times \hat r \!\right\} \) \cont \nonumber\\ & &\qquad\qquad\cont\qquad e^{-\frac{i\omega (r_0'+r-\hat r \cdot \vec r\,')}{c}}\Biggr |^2\label{eq:finRes}
\end{eqnarray}
\english{which, if we restrict ourselves to the special case $h=1$, means}
\begin{equation}
\frac{d^2W}{d\omega d\Omega}= \frac{1}{4\pi^2 c^3} \Biggl| \,\int_{\mathbb R^3} \mkern-8mu d^3 r'  \int_{ct_0-r+\hat r\cdot\vec r\,'}^{ct_1-r+\hat r\cdot\vec r\,'} \mkern-30mu dr'_{0}\; \hat r \times \ableit{}{r'_0} \vec \jmath\,(r'_0, \vec r\,')\times \hat r \;  e^{-\frac{i\omega}{c}( r'_0+r-\hat r \cdot \vec r\,')}  \Biggr|^2. \nopaper{\nonumber\\}\label{eq:qm_spec}
\end{equation}%
\english{This is still difficult to evaluate because of the variable bounds of integration. We expand them as needed and compensate the error in the integrand:}
\begin{eqnarray}
\frac{d^2W}{d\omega d\Omega}& \oveqzit{eq:finRes}& \frac{1}{4\pi^2c^3} \Biggl |\,    \int_{\mathbb R^3} \!\! d^3 r' \!\! \int_{c t_0'}^{ct_1'} \!\! dr_0' \, g(r_0',\vec r\,') \! \(\hat r \times \left\{ \ableit{}{r_0'}\vec \jmath\,(r_0', \vec r\,') \times \hat r \!\right\} \) \cont\nonumber\\ & &\qquad\qquad\cont\qquad e^{-\frac{i\omega (r_0'+r-\hat r \cdot \vec r\,')}{c}}\Biggr |^2 \label{eq:finRes2}\\
ct_0'&\ovdefeq&\underset{\forall  \vec r\,' \in \mathbb R^3}{min}({c t_0 - r + \hat r \cdot \vec r\,'})=const.\label{eq:limits1a}\\
ct_1'&\ovdefeq&\underset{\forall  \vec r\,' \in \mathbb R^3}{max}({c t_1 - r + \hat r \cdot \vec r\,'})=const.\label{eq:limits1b}\\
g(r_0',\vec r\,') &\ovdefeq& \sqrt{h_{\(\frac{r_0+ r - \hat r \cdot \vec r\,'}{c}\)}} \Theta(r_0'- c t_0+ r - \hat r \cdot \vec r\,')\Theta(-r_0'  + c t_1- r + \hat r \cdot \vec r\,')\nonumber\\*\label{eq:thetas}
 && \text{\english{where}\deutsch{wobei}~} \Theta(x)= \begin{cases}
     1     &\forall x>0 , \\
     0     & \text{\english{otherwise}\deutsch{sonst}}.
\end{cases} 
\end{eqnarray}
\english{\noindent Here, the time interval $t_0 \ldots t_1$ is the detection interval, and $t_0' \ldots t_1'$ are the times for which we have to supply the current data. Thus, the latter mark the start and end of our numerical simulation. 
In the exponent, $\frac{\omega r}{c}$ is large and constant, and can therefore be omitted. In fact, it should be omitted for numerical reasons: The functions sine and cosine do not work well for large arguments.
The inversion of \zit{eq:limits1a} and   \zit{eq:limits1b}  tells us the detection time limits:}
\begin{eqnarray}
ct_0&=&c t_0' + r -\underset{\forall  \vec r\,' \in \mathbb R^3}{min}({ \hat r \cdot \vec r\,'})\label{eq:limits2a}\\
ct_1&=&c t_1' + r -\underset{\forall  \vec r\,' \in \mathbb R^3}{max}({ \hat r \cdot \vec r\,'})\label{eq:limits2b}
\end{eqnarray}
\english{Let $V_{r_0}=supp(\vec \jmath\,(r_0, \vec r\,))$ be the support of the current at the time $r_0$ and $V= \underset{\forall r_0}{\bigcup} V_{r_0}$ the sum of all these. Then, in \zit{eq:finRes2}, it is possible to replace the spatial integral over $\mathbb R^3$  by another one over just $V$, and to calculate the times $t_0$ and $t_1$ in \zit{eq:limits2a} and \zit{eq:limits2b} as well (use $\forall  \vec r\,' \in V \text{ instead of } \mathbb R^3$) -- at least in principle. As we do not want to run our code twice -- once to exactly evaluate $V$, the second time to actually calculate the spectrum (to save all currents from the first run would definitely exceed any storage capacity) -- we simply choose $V$ based upon classical estimates of the particle trajectory as a spherical volume (which facilitates arbitrary observation directions) of radius $R$, which is large enough
to enclose all $V_{r_0}$. This is shown in \fig{fig:Skizze_Spektrum}. If the current is not localized at the origin, as in the above derivations, but at the middle $\vec r_K$ of the spherical volume $V$, then all positions vectors $\vec r, \vec r\,'$ have to be moved accordingly:}
\begin{equation}
\vec r \rightarrow \vec r - \vec r_K, \quad
\vec r\,' \rightarrow \vec r\,'  - \vec r_K,\quad
\hat r = \frac{\vec r}{|\vec r\,|} \rightarrow \frac{\vec r- \vec r_K}{|\vec r- \vec r_K|}, \quad 
r=|\vec r\,| \rightarrow  |\vec r- \vec r_K|\label{eq:verschiebe}
\end{equation}
\english{If $ \frac{\vec r- \vec r_K}{|\vec r- \vec r_K|}$ points from the mid point $\vec r_K$ towards the observer $\vec r$,  then one obtains from \zit{eq:limits2a} and  \zit{eq:limits2b}, taking into account \zit{eq:verschiebe} }
\begin{eqnarray}
ct_0&=&c t_0' +  |\vec r- \vec r_K| -\underset{\forall  \vec r\,' \in  V}{min}\!\(\frac{\vec r- \vec r_K}{|\vec r- \vec r_K|}  \cdot ( \vec r\,'  - \vec r_K)\)= c t_0' + |\vec r- \vec r_K| + R  \nonumber\\
ct_1&=&c t_1' + |\vec r- \vec r_K| -\underset{\forall  \vec r\,' \in  V}{max}\!\(\frac{\vec r- \vec r_K}{|\vec r- \vec r_K|}  \cdot ( \vec r\,'  - \vec r_K)\)= c t_1' +  |\vec r- \vec r_K|  - R \nonumber\\ \label{eq:limits3}
\end{eqnarray}
\english{Changes according to \zit{eq:verschiebe} have to be applied to all equations \zit{eq:finRes2} to \zit{eq:thetas}, of course.}
%
\english{The following explanation is meant to improve the readers intuitive understanding of the above purely mathematical description:
If we designate the edge of volume $V$ that lies furthest away from a given, fixed observer by the label a) and the other, closest edge by b), as shown in \fig{fig:Skizze_Spektrum}, then obviously the time that the radiation needs to travel from a) to the observer is longer than from b). The simulation starts at $r'_0=ct_0'$. One would be inclined to activate the detector at $t_0 =t_0'+[$travel time b) to the detector$]$ and to close it at $t_1=t_1'+[$travel time from a) to detector$]$ in order to avoid letting any radiation go by undetected. However, for consistency, then one would also have to take into account any radiation that reaches the detector at time $t_0$ as defined before, but which originates from position a) at a time before the beginning of our actual calculation, and about which therefore, no information is available. Thus, our conservative approach is to open the detector a little bit later. The choice  $t_0 =t_0'+[$travel time from a) to the detector$]$ throws away some radiation of origin b), but now everything that is detected is properly taken care of. Similarly, at time $t_1$ as defined above there would be a lack of any information about radiation that is coming from b) and superimposing the radiation that is still arriving from a). Therefore, the conservative choice  $t_1=t_1'+[$travel time from b) to the detector$]$ has to be made here, too.
Considering \equ{eq:limits3}, one easily notices that the detection interval can become rather short for large values of $R$. Thus, a spatially extended motion can only be analyzed if it is slow, or more exactly speaking, if $t_0'$ and $t_1'$ are widely separated in time. At least for a particle initially at rest, $t'_0$ can be moved into the past as much as required without altering $R$, whenever this is not the case.}
%
%
%
\subsection{Application to quantum mechanics}
\label{sec:qmspec}
\english{In relativistic quantum mechanics, the probability current density is given in the following way \cite{Bjorken_Drell}}
\begin{equation}
j_{prob}^\mu = c\psi^{\dagger}\gamma^0\gamma^\mu\psi, \quad{\rm where}\quad\gamma^0=\beta \quad{\rm and}\quad \vec\gamma=\beta\vec\alpha,
\end{equation}
\english{and can be easily decomposed into the vectorial current density}
\begin{equation}
\vec \jmath_{prob} = c \psi^\dagger\vec\alpha\psi \label{eq:current_density_Z8}
\end{equation}
\english{and the scalar density}
\begin{equation}
\rho_{prob} = \psi^\dagger \psi \label{eq:density_Z8}.
\end{equation}
%
\english{For the charge current $\vec \jmath$ in the context of Dirac theory, we choose}%
\onlypaper{\begin{equation}
\vec \jmath = q \, \vec \jmath_{prob} = q c\, \psi^\dagger\vec\alpha\psi  \label{eq:chargecurrent}
\end{equation}}
\english{This choice is less obvious than it seems. For a extensive discussion of this matter, see \cite{Gurtler:1974}.  The time derivative} $\ableit{}{r_0} \vec \jmath=\frac{1}{c}\ableit{}{t} \vec \jmath$ \english{is derived in the following way:}
\onlypaper{\begin{eqnarray}
\ableit{}{t} \vec \jmath &\oveqzit{eq:chargecurrent}& q c\, \ableit{\psi^\dagger}{t}\vec\alpha\psi  +q c\, \psi^\dagger\vec\alpha\ableit{\psi}{t}  \label{eq:dtj}\\
\ableit{\psi}{t}&\oveqzit{eq:Dirac_Hestenes}& \frac{1}{ i\hbar}\(  c\vec\alpha\cdot \(\frac{\hbar}{i}\vec\nabla-\frac{q}{c}\vec A\) \psi + \beta mc^2\psi +q A_0 \psi\) \label{eq:dtPsi}\\
\frac{\hbar}{i}\vec\nabla  \psi  &=& {\Fourier^{-1}}\; \hbar \vec k\;{\Fourier} \psi. \label{eq:momOp}
\end{eqnarray}}

\onlypaper{\english{Note that \zit{eq:dtPsi} is an immediate result from the Dirac equation  \zit{eq:Dirac_Hestenes}.}}
\english{The momentum operator \zit{eq:momOp} is transformed by means of Fourier transforms into a simple multiplication operator.}
\subsection{Reduction to two dimensions}
\english{We have to reduce  \zit{eq:finRes2} by means of \zit{eq:2dAnsatz} to the two-dimensional case: If the observer is located inside the plane, i.e. $\hat r \cdot \vec r\,' = \hat r \cdot \vec r\,'_\perp$ then  equations \zit{eq:finRes2},\zit{eq:limits1a},\zit{eq:limits1b} and \zit{eq:thetas} are transformed into}
\begin{eqnarray}
\frac{d^2W}{d\omega d\Omega}& \oveqzit{eq:finRes2}&  \frac{1}{4\pi^2c^3} \Biggl |\,    \int\!\! d^2 r'_\perp \mkern-10mu \int_{c t_0'}^{ct_1'}\!\! dr_0' \, g(r_0',\vec r\,'_\perp) \! \(\hat r \times \left\{  \int\!\! dr'_\parallel\ableit{}{r_0'}\vec \jmath\,(r_0', \vec r\,'_\perp) \times \hat r \!\right\} \)  \cont\nonumber\\ & &\qquad\qquad\cont\qquad e^{-\frac{i\omega (r_0'+r-\hat r \cdot \vec r\,'_\perp)}{c}}\Biggr |^2 \label{eq:finRes3}\\
ct_0'& \oveqzit{eq:limits1a}&\underset{\forall  \vec r\,'_\perp}{min}({c t_0 - r + \hat r \cdot \vec r\,'_\perp})\label{eq:limits3a}\\
ct_1'& \oveqzit{eq:limits1b}&\underset{\forall  \vec r\,'_\perp}{max}({c t_1 - r + \hat r \cdot \vec r\,'_\perp})\label{eq:limits3b}\\
g(r_0',\vec r\,'_\perp) &\oveqzit{eq:thetas}& \sqrt{h_{\(\!\frac{r_0+ r - \hat r \cdot \vec r\,'_\perp}{c}\!\)}} \Theta(r_0'- c t_0+ r - \hat r \cdot \vec r\,'_\perp)\qquad\cont \nonumber\\ &&\cont\qquad\Theta(-r_0'  + c t_1- r + \hat r \cdot \vec r\,'_\perp) \label{eq:thetas3}
\end{eqnarray}
\english{To evaluate the above, we need}
\onlypaper{\begin{eqnarray}
 \int\!\! dr'_\parallel \ableit{}{r_0} \vec \jmath &\oveqzit{eq:dtj}& q   \int\!\! dr'_\parallel \(\ableit{\psi^\dagger}{t}\vec\alpha\psi  +  \psi^\dagger\vec\alpha\ableit{\psi}{t}\) \nonumber\\
&=&  q \( \left[ \int\!\! dr'_\parallel  \psi^\dagger\vec\alpha\ableit{\psi}{t}   \right]^\dagger + \left[ \int\!\! dr'_\parallel   \psi^\dagger\vec\alpha\ableit{\psi}{t}   \right]\)\label{eq:intj},
\end{eqnarray}}
\english{and as a part of the above}
\onlypaper{\begin{eqnarray}
 \int\!\! dr'_\parallel \,  \psi^\dagger\vec\alpha\ableit{\psi}{t} \quad &\ovuneqzit{eq:2dAnsatz}{eq:Dirac2D}& \quad\psi'^\dagger\vec\alpha  \( \frac{1}{ i\hbar}\left\{\!  c\vec\alpha\cdot \(\frac{\hbar}{i}\vec\nabla_\perp 
 -\frac{q}{c}\vec A\)  + \beta mc^2 +q A_0\!\right\} \psi'   \)\cont\nonumber\\&& \quad\cont\quad \underbrace{\int\!\! dr'_\parallel \chi  \chi^\dagger }_{=\hbar\,\delta(0 )} \nonumber\\
&=&\quad \psi'^\dagger\vec\alpha  \biggl [ \frac{1}{ i\hbar} \bigl\{  \Fourier_\perp^{-1}c\vec\alpha\cdot \( \hbar\vec k_\perp
\)\Fourier_\perp \psi' + \nonumber\\*
&&\quad\qquad+\(\beta mc^2  +q A_0-q \(\vec\alpha\cdot \vec A\,\) \)\psi' \bigr\}  \biggr ]  \hbar\,\delta(0)   \label{eq:intpsi}.
\end{eqnarray}}
\english{Finally, insert equation \zit{eq:intpsi} (apart from the factor $ \hbar\delta(0)$, which has to be omitted, cf. appendix \ref{sec:Dirac_Hestenes}) into \zit{eq:intj} and the result into \zit{eq:finRes3}. Together with \zit{eq:limits3a}, \zit{eq:limits3b} and  \zit{eq:thetas3} this would give (for the case $h=1$) the final formula used in our code. Because of its size, we do not print it again.}
\subsection{Numerical limitations}
\label{sec:qm_spectrum_limitations}
\english{Tests showed that the derivative of the current as shown before in equations \zit{eq:intj} and \zit{eq:intpsi}, although correct in theory, cannot be numerically calculated correctly this way. }%
\english{Apparently, the reason is to be found in  \equ{eq:dtPsi}: This equation is only true, if $\psi$ is a exact solution of the Dirac equation. Small errors in, for example, the phase of the wave function can have a large influence on the result. However, the current  \zit{eq:chargecurrent} itself is independent of the phase. Therefore, a less elegant, but numerically superior method is a finite difference scheme like the following:}%
\begin{equation}
\ableit{}{t} \vec \jmath=   \frac{\vec \jmath\,(t+\Delta t) -\vec \jmath\,(t)}{\Delta t} \quad \text{ \deutsch{wo}\english{where} }\quad  \vec \jmath\,(t)\oveqzit{eq:chargecurrent} q c\, \psi(t)^\dagger\vec\alpha\psi(t)
\end{equation}%
\english{A higher order scheme would be desirable, however this would require a rather impracticable multiple storage of the wave function. Fortunately, $\Delta t$ has to be chosen only small enough as to let $\omega_{max}=\frac{\pi}{\Delta t}$ cover the whole spectral range that is to be expected. It can therefore be much larger than the step size of the numerical propagation of $\psi$.}
\subsection{Non-relativistic limit}
\label{sec:qm_spectrum_nonrel}
\english{Just to gain some trust into the above derivation, we now look at its non-relativistic and spin-less limit. It can be obtained by ignoring small retardation effects $\hat r \cdot \vec r\,'\approx 0$ and using the Schr\"odinger charge current\footnote{The non-relativistic limit for a particle with spin could be obtained from the Pauli theory by insertion of the Pauli charge current, which differs from \zit{eq:SchroedingerLadungsstrom} by an extra spin-dependent term \cite{Gurtler:1974}.} (which depends on the Schr\"odinger wave function $\psi_t\in \mathbb C$)}
\begin{equation}
\vec \jmath = q\left\{ \frac{\hbar}{2 i m}(\psi_t^\dagger (\vec \nabla \psi_t) - ( \vec \nabla \psi_t )^\dagger \psi_t)-\frac{q}{mc}\vec A \psi_t^\dagger\psi_t \right\}.\label{eq:SchroedingerLadungsstrom}
\end{equation}
\english{Similar to the classical theory's limit in \autoref{sec:cm_spectrum_nonrel}, we insert  $\hat r \cdot \vec r\,'\approx 0$ and \zit{eq:SchroedingerLadungsstrom}  into \zit{eq:qm_spec} and, after some work, arrive at:}
\begin{equation}
\frac{d^2W}{d\omega d\Omega} \ovuneqzit{eq:qm_spec}{eq:SchroedingerLadungsstrom} \frac{q^2}{4\pi^2 c^5} \Biggl|  \;\int_{ct_0-r}^{ct_1-r} \mkern-10mu dr'_{0}\; \hat r \times \(\frac{1}{m} \ableit{}{t'}\(\psi_{t'}, \vec p \psi_{t'}\)\)\times \hat r \;  e^{-\frac{i\omega}{c}r'_0}  \Biggr|^2.\label{eq:ZE_Impulsintegral}
\end{equation}
\english{Here $r'_0= ct'$ and $\vec p$ designates the momentum operator}
\begin{equation}
\vec p \psi_t= -i\hbar \vec \nabla\psi_t - \frac{q}{c}\vec A \psi_t,
\end{equation}
\english{whose expectation value is defined, for the case of perfect conservation of normalization $\(\psi_t, \psi_t\)=1 \quad \forall t$, as follows:}
\begin{equation}
\(\psi_{t'}, \vec p \psi_{t'}\) = \langle \vec p \;\rangle_{\psi_{t'}} \(\psi_{t'}, \psi_{t'}\) =  \langle \vec p \;\rangle_{\psi_{t'}}.\label{eq:EWWimpuls}
\end{equation}
\english{At this point, we can  employ the Ehrenfest theorem and, for the case of a vanishing magnetic field $\vec B\approx \vec 0$, we write}
\begin{equation}
 \ableit{}{t}\langle\vec p \,\rangle_{\psi_t} \ovuneq{\text{Ehrenfest}}{\text{}}  \langle q \vec E(\vec x,t)\rangle_{\psi_t}\label{eq:EhrenfestNewton}
\end{equation}
\english{to arrive at a simple formula:
}
{\begin{equation}
\frac{d^2W}{d\omega d\Omega}\ovuneq{\zit{eq:ZE_Impulsintegral}}{\zit{eq:EWWimpuls},\zit{eq:EhrenfestNewton}} \frac{q^2}{4\pi^2 c^5} \Biggl|  \;\int_{ct_0-r}^{ct_1-r} \mkern-10mu dr'_{0}\; \hat r \times   \left\langle\frac{q}{m}\vec E (\vec x,t')\right\rangle_{\psi_{t'}}  \times \hat r \;  e^{-\frac{i\omega}{c}r'_0}  \Biggr|^2.\label{eq:ZE_E_integral}
\end{equation}}%
\english{Calling $\vec a\ovdefeq  \frac{q}{m}\vec E $ and $ \langle\vec a \rangle_{\perp(t')} \ovdefeq \hat r \times \langle \vec a  \rangle_{\psi_t'}\times \hat r$ we obtain }
\begin{eqnarray}
\frac{d^2W}{d\omega d\Omega}&=& \frac{q^2}{4\pi^2 c^5} \Biggl|  \;\int_{ct_0-r}^{ct_1-r} \mkern-10mu dr'_{0}\;   \langle \vec a  \rangle_{\perp (t')} \;  e^{-\frac{i\omega}{c}r'_0}  \Biggr|^2\\
&=& \frac{q^2}{2\pi  c^3} \Biggl| \frac{1}{\sqrt{2\pi}} \;\int_{t_0-r/c}^{t_1-r/c} \mkern-10mu dt' \;   \langle \vec a  \rangle_{\perp (t')} \;  e^{-i\omega t'}  \Biggr|^2\label{eq:BeschlIntegral} \quad\ovun{t_0\rightarrow -\infty}{t_1\rightarrow +\infty}{\longrightarrow}   \quad\frac{q^2}{2\pi  c^3} \Biggl|  \langle \vec a  \rangle_{\perp(\omega)} \Biggr|^2.\label{eq:d2Wqm_nonrel}
\end{eqnarray}
\english{If we expand the integration interval to infinity, as indicated above, and replace $\vec a_\perp \leftrightarrow \langle \vec a  \rangle_{\perp}$, then we recover the classical result \zit{eq:d2Wclassical_nonrel} in  \zit{eq:d2Wqm_nonrel}.  It returns the spectrum as proportional to the Fourier transform of the perpendicular component of the mean acceleration.\footnote{Note that most authors ignore the correct prefactor and give the spectrum in ``arbitrary units'' only.} If however, we neglect the retardation $r$ in the integration limits of  \equ{eq:BeschlIntegral}, then we obtain the result from  \cite{Burnett:1992}.\footnote{The authors of \cite{Burnett:1992} employ the length gauge $\vec E(t)= -\vec \nabla A_0$ with $A_0=  -\vec E(t) \cdot \vec x $ and the dipole approximation $\vec A=\vec 0$.}}
 
\section{Spectra of free electrons in a laser field}\label{chap:results_spectrum}
\english{In order to appreciate the work carried out in \autoref{sec:qm_rad}, we have made a comparative study of five different methods. In this study, the physical parameters are primarily chosen in order to magnify the differences in the spectra and to speed up calculations.}

\english{The scenario is the following one: A free electronic wave packet is propagated over 20 cycles, two of which are dedicated to the field amplitude's $\sin^2$-shaped turn-on and turn-off phase, respectively. In order to see a reasonable number of harmonics in the spectrum of an otherwise free electron  \cite{Vachaspati:1962}, such a large number of cycles is necessary. A free electron is favourable to the numerics: There are, at least in polarization direction, no large canonical momentum components, and the wave packet is not torn apart by scattering in addition to its unavoidable natural spreading. Therefore, it is possible to use small grids of rather low spatial resolution. The large number of cycles implies a high frequency in order to keep the overall propagation time short. Note that the time resolution for Dirac calculations is almost independent of the problem\footnote{Extremely high kinetic energies in the MeV regime are an exception to this rule, see \cite{Mocken:2003c}. They require an even higher temporal resolution since $E=\gamma mc^2$ and $\gamma > 1$.} and only given by  $\Delta t \lesssim \frac{\hbar}{E}\approx\frac{\hbar}{mc^2}$. Therefore the frequency was chosen as $\omega=7.120$~a.u. (corresponding to $6.4$~nm).\footnote{This particular frequency is the highest one that is indicated for the future FEL of the so-called ,,TESLA-Test-Facility (Phase 2)'' at DESY in Hamburg \cite{Tesla:2001}.} In order to stay in the strongly relativistic regime -- it is only here that we can expect appreciable  differences -- the intensity has to be set to the challenging value  $E_0=300$~a.u. ($I=3.15\times10^{21}$~W/cm$^2$).
The following calculations were carried out:}
\begin{itemize}
\item \english{the classically-relativistic trajectory of a point particle according to the procedure described in  \autoref{chap:rk4}.}
\item \english{the quantum mechanical propagation of a Gaussian wave packet according to the Dirac equation as shown in  \cite{Mocken:2003b}.}
\item \english{the quantum mechanical propagation of a Gaussian wave packet according to the Schr\"odinger equation.\footnote{This calculation was contributed by Andreas Staudt, who is the author of the required Schr\"odinger-code.}}
\end{itemize}
\english{From these data sets, the following spectra were produced:}
\begin{itemize}
\item \english{Based on the classical trajectory data, the following was calculated:}
\begin{itemize}
\item \english{the relativistically correct spectrum according to \equ{eq:d2Wclassical}. This will be labelled ``classical theory (exact)'' in the figures to follow.}
\item  \english{the non-relativistically approximated spectrum according to  \equ{eq:d2Wclassical_nonrel}, labelled ``classical theory (approx.)''.}
\end{itemize}
\english{In both cases we used the same (relativistically correct) trajectory data set, i.e. the distinction relativistic vs. non-relativistic only refers to the way that the spectrum is evaluated.}
\end{itemize}
\begin{itemize}
\item \english{From the single data set according to the Dirac theory, two different kinds of spectra were extracted:}
\begin{itemize}
\item \english{the fully relativistic one (``Dirac'') according to the method presented in  subsections \ref{sec:qm_spectrum} to \ref{sec:qm_spectrum_limitations}: It contains no non-relativistic approximations and takes into account all retardation and spin effects.}
\item \english{a half-relativistic one  (``Dirac-Schr\"odinger''): the expectation value of the electric field was evaluated from the density distribution, and then, according to the procedure in  \autoref{sec:qm_spectrum_nonrel}, the spectrum was calculated. In other words, the relativistically correct Dirac probability density was fed into a non-relativistic procedure, which was originally developed for the Schr\"odinger probability density. Spin effects are taken care of for the propagation, but not for the spectrum calculation, i.e. for the charge current.}
\end{itemize}
\item \english{From Schr\"odinger data set (``Schr\"odinger''), the spectrum was evaluated according to the method in \autoref{sec:qm_spectrum_nonrel}. This represents an entirely non-relativistic version, with no spin effects at all.}
\end{itemize}
\english{Additionally, the observation angle $\vartheta$ is varied from $0$° to $90$° in steps of $30$°. The ``Dirac-Schr\"odinger'' method allows us to differentiate between effects caused by different wave function propagation methods and those that originate from the different treatment of the emitted radiation.}
\subsection{Comparison of the results}
\begin{figure}[p]
\begin{center}
	\includegraphics[width=\textwidth]{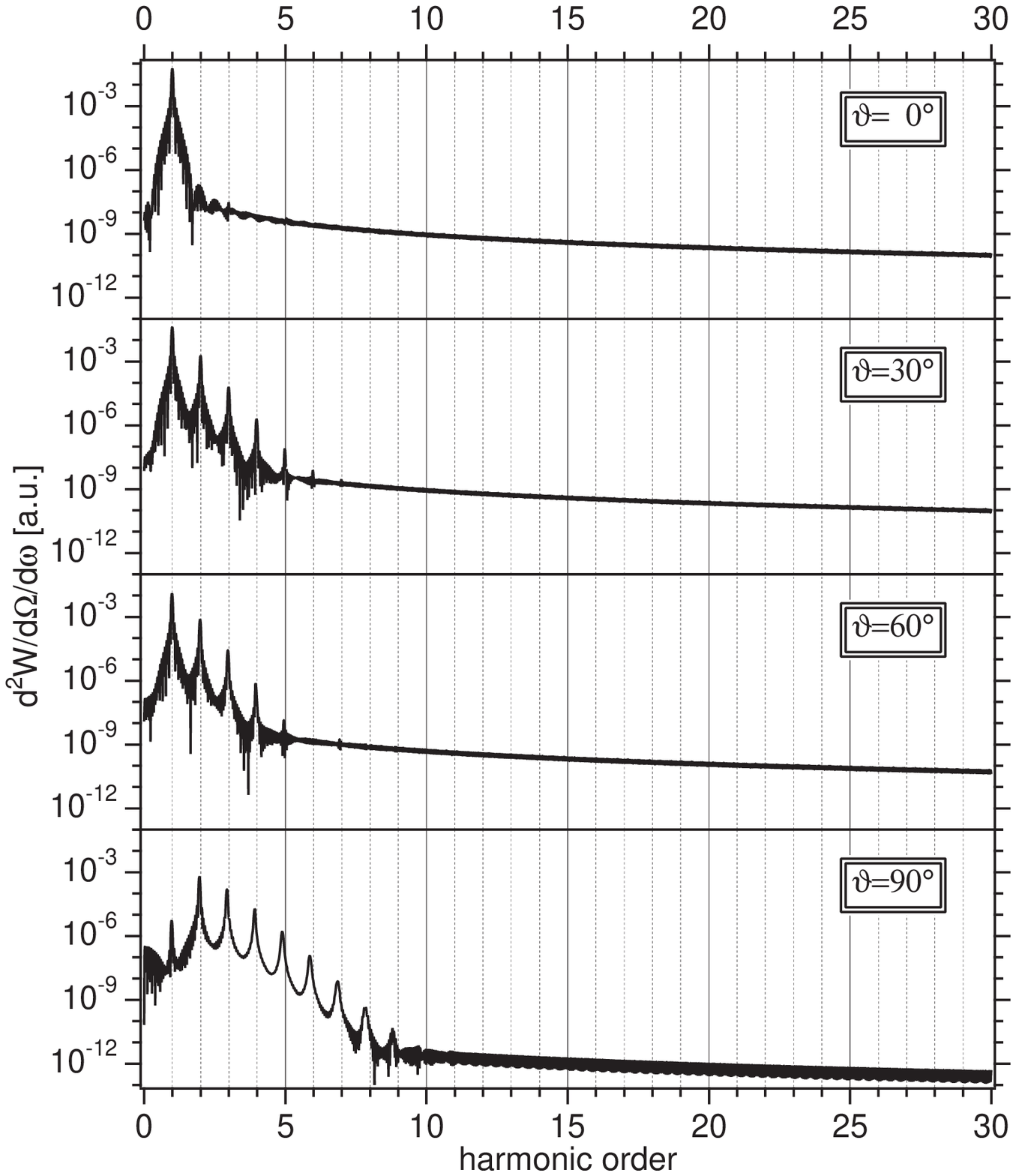}
\caption[Spectrum angle dependence]{\english{Spectra according to the Dirac calculation: The plot shows the spectrum, evaluated according to the method from  subsections \ref{sec:qm_spectrum} to \ref{sec:qm_spectrum_limitations}, of a free electron that interacts with a laser pulse of 20 cycles (two of which are reserved for turn-on and turn-off) with an amplitude $E_0=300$~a.u. ($I=3.15\times10^{21}~$W/cm$^2$) and frequency $\omega=7.12$~a.u. ($\lambda=6.4$~nm). The fully relativistic spectrum due to the two-dimensional Dirac charge current is shown for several different angles of observation  $\vartheta$ with respect to the propagation direction.}}
\label{fig:Dirac_Spektren}
\end{center}
\end{figure}
\begin{figure}[p]
\begin{center}
	\includegraphics[width=\textwidth]{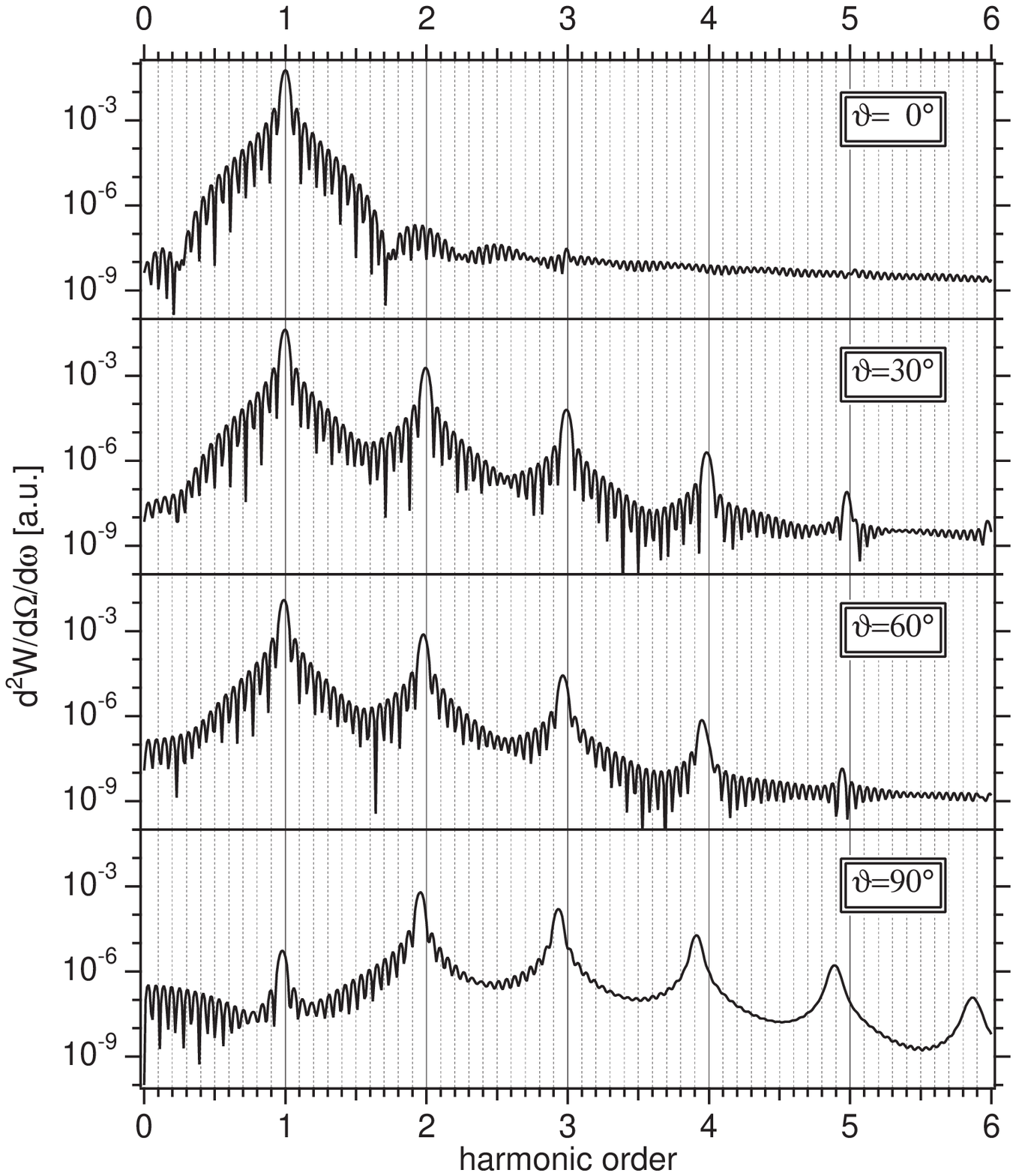}
\caption{\english{Spectra according to the Dirac calculation; zoom of  \fig{fig:Dirac_Spektren}: The appearance of the even orders for non-zero angles and the increase of both intensity and red shift of all  harmonics with growing observation angle is clearly visible. At $\vartheta=90$°, the fundamental line is suppressed.}}
\label{fig:Dirac_Spektren_zoom}
\end{center}
\end{figure}
%
%
\english{We point out the following features of the spectrum that was calculated using the Dirac charge current in  \fig{fig:Dirac_Spektren} and the magnification thereof in \fig{fig:Dirac_Spektren_zoom}:
For  $\vartheta=0$°, there are only harmonics of first, third and fifth order. Third and fifth order are very weak and hard to distinguish from the oscillating underground. No red shift is found.
At  $\vartheta=30$°, there are more harmonic orders: Up to the seventh order both even and odd harmonics can be seen, and a small red shift is found. This effect is enlarged for $\vartheta=60$°. For $\vartheta=90$° we find the largest red shift and harmonics of up to ninth order. The intensity of the fundamental line decreases to a level even below the one of the second to fourth order harmonics. The underground intensity also decreases.}
\begin{figure}[p]
\begin{center}
	\includegraphics[width=\textwidth]{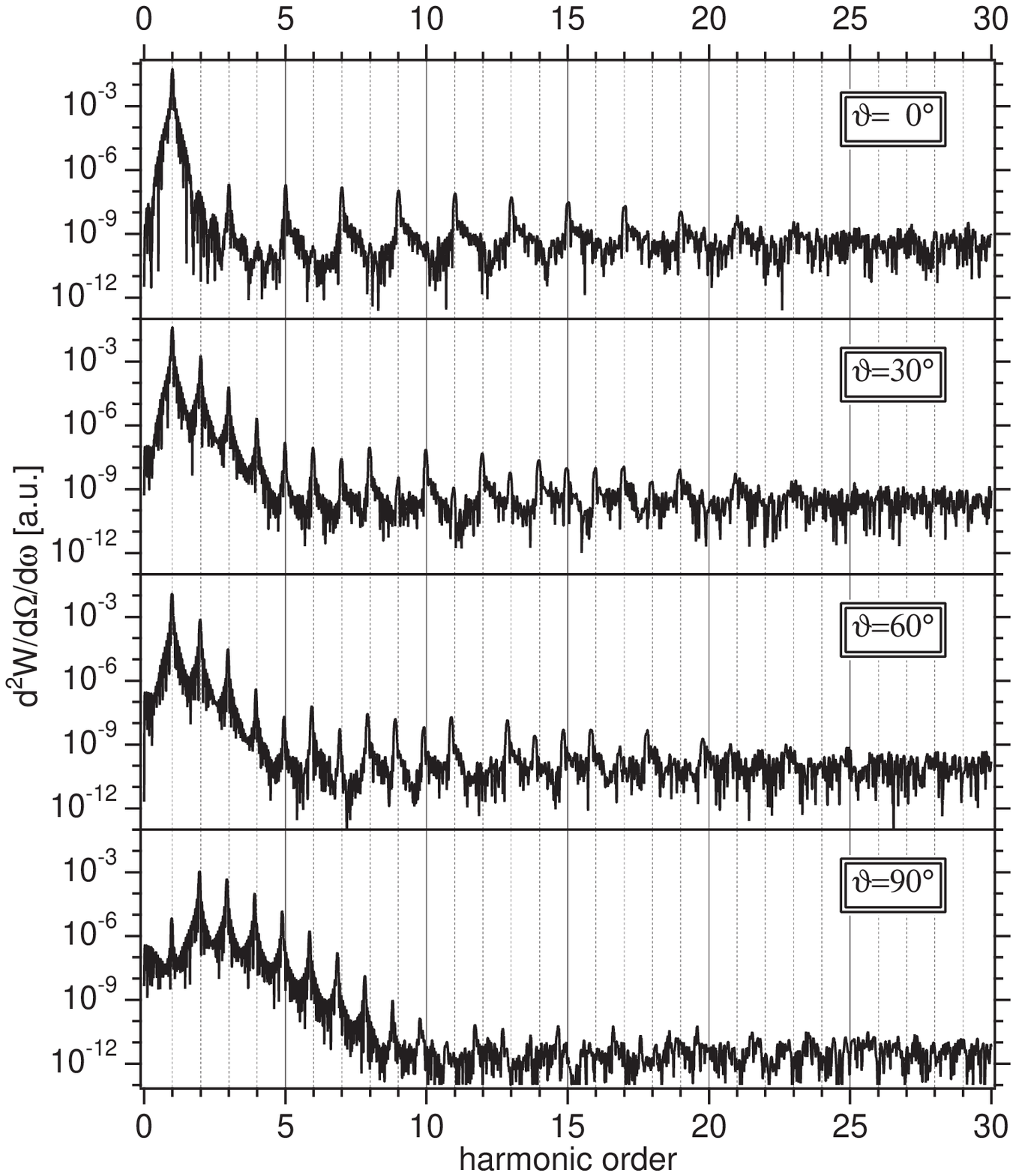}
\caption{\english{Spectrum according to the classical calculation: The plot shows the spectrum of a free electron in interaction with a 20 cycle laser pulse (including two cycles for turn-on and turn-off, respectively) with an amplitude of $E_0=300$~a.u. ($I=3.15\times10^{21}$~W/cm$^2$) and a frequency of $\omega=7.12$~a.u. ($\lambda=6.4$~nm). The fully relativistically calculated spectrum due to two-dimensional motion of a classical point particle is shown for several different angles of observation  $\vartheta$  with respect to the propagation direction.}}
\label{fig:Klassik_Spektren}
\end{center}
\end{figure}
\begin{figure}[p]
\begin{center}
	\includegraphics[width=\textwidth]{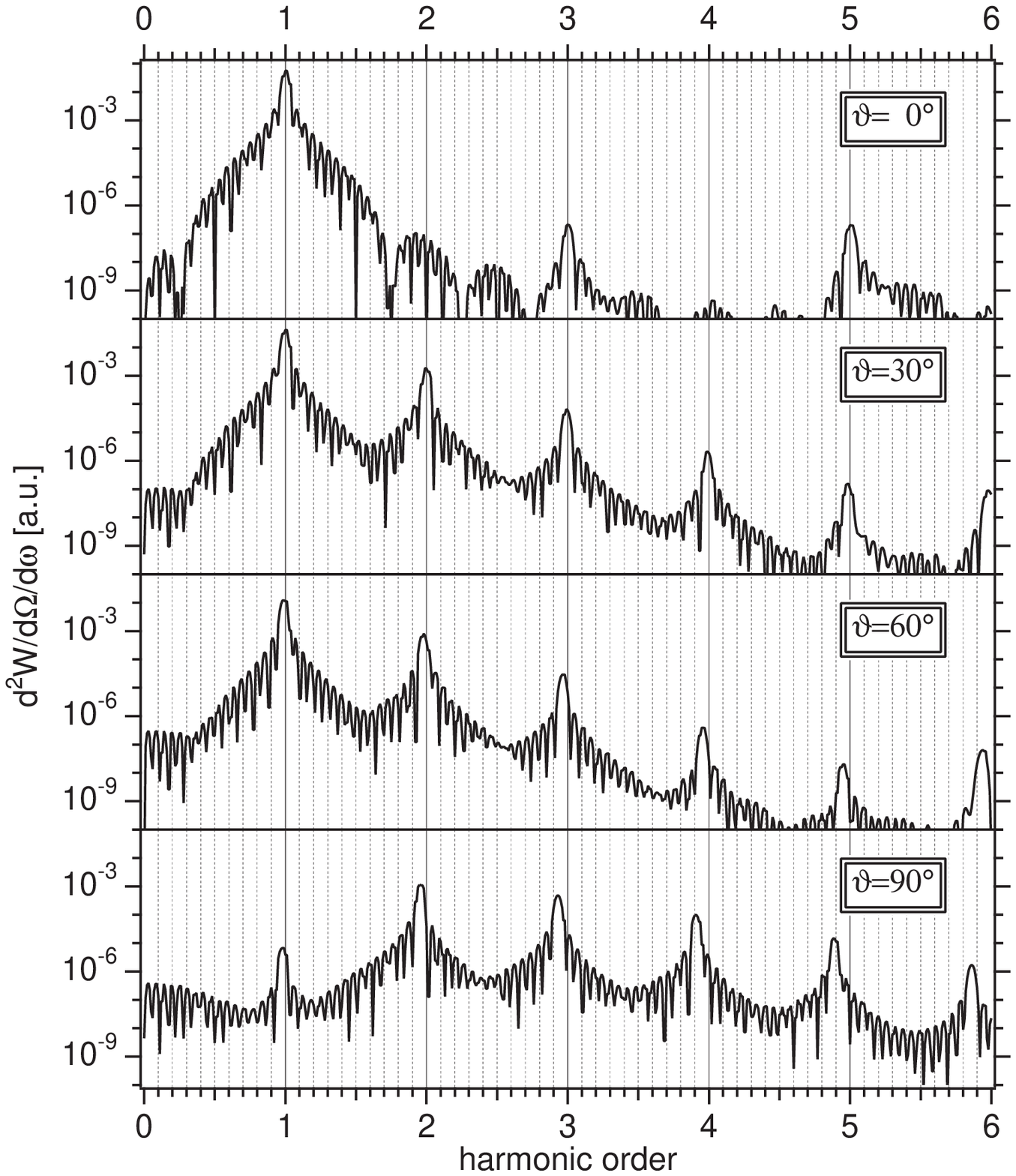}
\caption{\english{Spectrum according to the classical calculation; zoom of \fig{fig:Klassik_Spektren}: The appearance of the even harmonic orders for non-zero angles and the increasing red shift of all harmonics with growing observation angle are easily visible. At $\vartheta=90$°, the fundamental line is suppressed.}}
\label{fig:Klassik_Spektren_zoom}
\end{center}
\end{figure}
\english{The underlying reasons are best understood by looking at the similar, but more intuitive classical picture. Looking at  the spectrum in \fig{fig:Klassik_Spektren} and its magnified version \fig{fig:Klassik_Spektren_zoom}, which are based upon the classical trajectory and the corresponding relativistically correct radiation formula, we notice that for $\vartheta=0$° there are only odd harmonics, but both even and odd ones up to about 19th order for all other angles. For order above $20$, the peaks disappear in the noise.
There is no red shift at $\vartheta=0$°, but an increasing one for larger angles. At $\vartheta=90$°, the intensity of the fundamental line drops below the one of the second to fifth order harmonic. The similarities between this classical calculation and the Dirac based one are obvious. The main difference is that due to the smeared out charge distribution in the quantum mechanical Dirac calculation, some sharp structures of the classical spectrum are averaged out, especially its higher frequency part.

The growth of the red shift with increasing angle can be understood qualitatively quite easily (cf. \cite{Keitel:1994}): In the rest frame of the electron, the laser frequency is red shifted because of the forward drift of the electron in the laser field, and the same is true for the oscillation frequency of the electron in polarization direction due to the laser's electric field component. If the electron drift is directed exactly towards the observer ($\vartheta=0$°), the latter sees this actually red-shifted frequency in the electron's radiation, which however is Doppler-shifted back to exactly its original position in the frequency domain. If the observation angle is larger ($\vartheta>0$°), the original red shift stays the same, but the compensating Doppler blue-shift decreases with increasing angle, leaving behind an increasing net red shift.}%

\english{The appearance of even and odd harmonics can be understood by the means of a symmetry consideration. One needs to consider the \nameequ{approximative formula}{eq:d2Wclassical_nonrel}, i.e. the magnitude squared of the Fourier transform of the acceleration component $\vec a_{\perp}$ perpendicular to the direction of observation. Reference \cite{Salamin:1996} provides the classical analytical solution for the particle velocity  $\vec \beta(\eta)$ in terms of the laser phase  $\eta$.  If we insert a simple harmonic cw-laser field in those formulas and take the time derivative, we obtain  for the acceleration something like
\begin{eqnarray}
\vec a = c \ableit{}{t}\vec \beta(\eta) = u(\eta) \sig{1} + g(\eta)\sig{2}
\end{eqnarray}
where  $u(\eta)$ is a rather complicated 
function, with odd symmetry with respect to both $\eta=0$ und $\eta=\frac{\pi}{2}$, whereas $g(\eta)$ features even symmetry with respect to $\eta=0$ and odd symmetry with respect to $\eta=\frac{\pi}{2}$. In the above, $\sig{1}$ and $\sig{2}$ are unit vectors in propagation and polarization direction, respectively. The properties of the Fourier transform then automatically lead to even harmonics for the $\sig{1}$-component of acceleration and odd ones for its $\sig{2}$-component. The latter ones are exclusively visible for the observation direction designated by $\vartheta=0$°, and only the former ones for  $\vartheta=90$°. At any angle in between, one would expect the appropriate mixture of both even and odd harmonic orders. 
The above analytical discussion explains the complete disappearance of even orders in the numerical calculations for $\vartheta=0$° just as well as the suppression of the fundamental (which is odd by itself) at $\vartheta=90$°. All deviations from the mathematically ideal behaviour have to be attributed to the relativistic corrections, the retardation  $r-\hat r\cdot \vec r\,'\ne 0$ and last but not least the usage of a time-limited, pulse-shaped laser pulse.}%

\english{Summing up the above, we can say that the classical and the quantum mechanical description are in good agreement and both deliver well understood spectra. One might therefore ask immediately for the motivation of the computationally expensive Dirac calculations. In fact, this particular scenario was set up mainly because we wanted to verify the correctness of the procedure described in subsections \ref{sec:qm_spectrum} to \ref{sec:qm_spectrum_limitations} using a system from which one would not expect large quantum effects, so any deviations from the well-known classical behaviour would easily reveals errors in the method or the code. The agreement between classical and quantum theory is likely to disappear however for laser-driven bound systems. The parameter regime usually referred to as ``tunnelling regime'', for which a description of the ``High Harmonic Generation''  process according to the recollision model \cite{Lewenstein:1994} is applicable, would be of high interest. Unfortunately, this regime is characterized by low frequencies and therefore long computational runtimes, which renders it impossible to study with our Dirac code, at least when simultaneously employing field strengths that are large enough to render relativistic and spin-related effects visible. Low intensities would be feasible, but not very interesting, as there are already a number of completely sufficient Schr\"odinger  \cite{Prager:2001}  and  Pauli \cite{KeitelHu:2001,Hu:2001} calculations for this regime.}
\begin{figure}[p]
\begin{center}
	\includegraphics[trim= 0 8 0 10, width=\textwidth,clip]{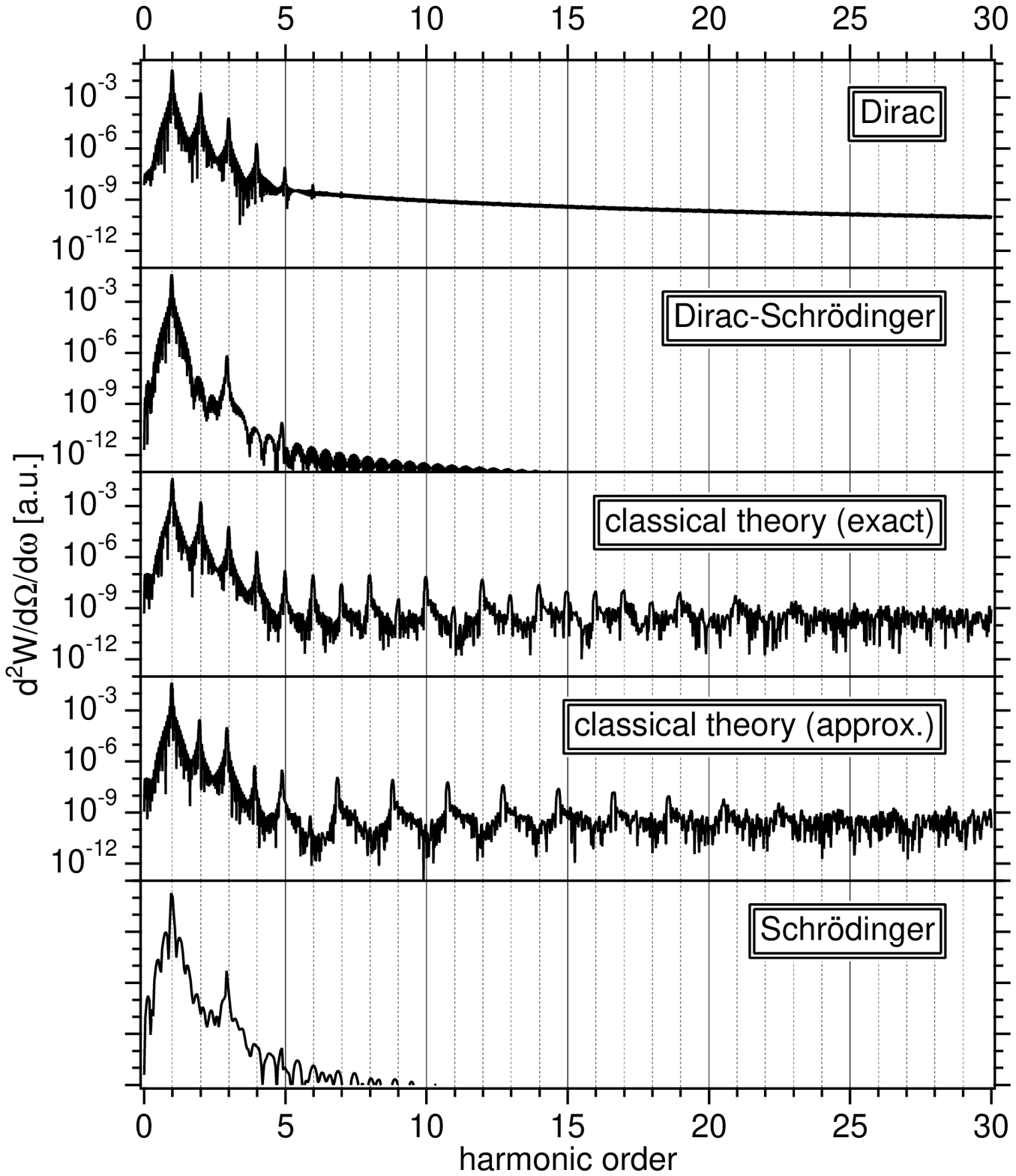}
\caption[Comparison]{\english{Comparison of several methods to evaluate the radiation spectrum:  The top graph is based on the method from subsections \ref{sec:qm_spectrum} to \ref{sec:qm_spectrum_limitations}  and shows the fully relativistic spectrum due to a two-dimensional Dirac charge current. 
The second graph  from top to bottom displays the spectrum obtained by Fourier transforming the perpendicular acceleration component $\vec a_{\perp}$, where the acceleration is taken as the expectation value of  $\frac{q}{m}\vec E(\vec x, t)$ with respect to the Dirac wave function, similar to the way shown in \autoref{sec:qm_spectrum_nonrel} for a Schrödinger wave function.
The third graph  shows the fully relativistic spectrum calculated for the classical trajectory according to \autoref{sec:cm_spectrum}, and the  fourth graph the one that can be obtained from the same trajectory data in the non-relativistic limit according to \autoref{sec:cm_spectrum_nonrel}. The bottom graph displays the result according to the method in \autoref{sec:qm_spectrum_nonrel} for a non-relativistic Schr\"odinger calculation. No vertical scale is given for this last calculation, which is a contribution of Andreas Staudt, because the prefactor is not known.}}
\label{fig:Vergleich}
\end{center}
\end{figure}
\begin{figure}[p]
\begin{center}
	\includegraphics[width=\textwidth]{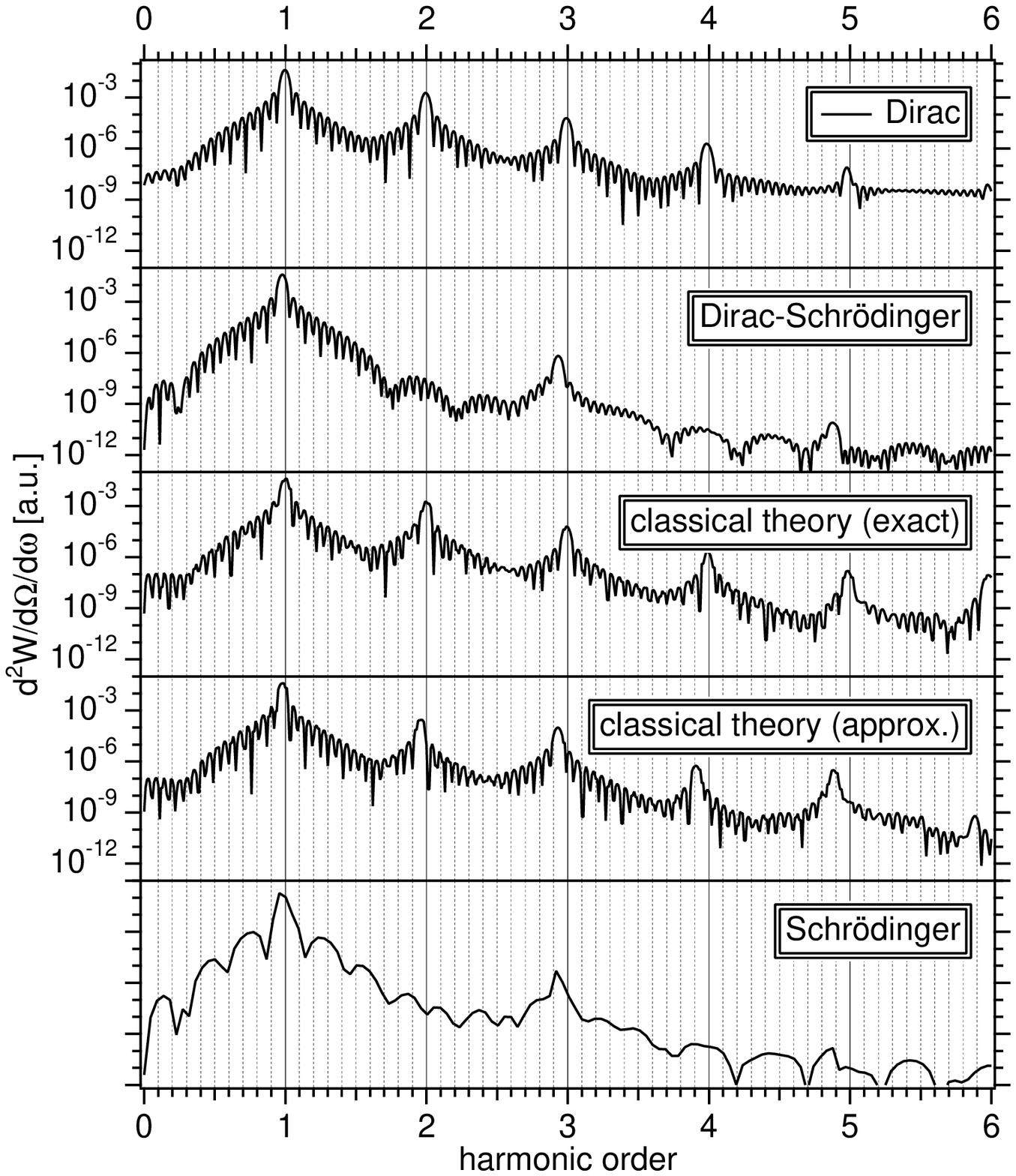}
\caption[zoom comparison]{\english{Enlarged view of \fig{fig:Vergleich}: The differences of the distinct methods in terms of red shift and harmonic orders are obvious. All methods that contain non-relativistic elements (Dirac-Schr\"odinger, approx. classical theory and Schr\"odinger) feature significant red shifts at the observation angle displayed here ($30$°). For the exact  classical  and the  Dirac-calculation noticeable red shifts only occur for much larger angles. Apart from that, even harmonic orders are missing in both the Dirac-Schr\"odinger and the Schr\"odinger calculation. The latter only seems to have less numerical noise than all the others, but this is due to a coarser frequency resolution.}}
\label{fig:Vergleich_zoom}
\end{center}
\end{figure}

\english{However, we note some interesting discrepancies when we compare the method from subsections \ref{sec:qm_spectrum} to \ref{sec:qm_spectrum_limitations} and the ordinary procedure as shown in \autoref{sec:qm_spectrum_nonrel}. The latter, or more exactly  \equ{eq:ZE_E_integral}, can, although originally deduced from Schr\"odinger theory, also be evaluated using the Dirac probability density: This is achieved by Fourier transforming the perpendicular acceleration component  $\vec a_{\perp}$ as in \equ{eq:d2Wqm_nonrel}, where the acceleration is taken as the expectation value of $\frac{q}{m}\vec E(\vec x, t)$  with respect to the Dirac wave function.
As an example, this is shown in \fig{fig:Vergleich} for an observation angle $\vartheta=30$° and enlarged in  \fig{fig:Vergleich_zoom} (second graph from top to bottom in both figures) side by side with the exact Dirac procedure, the also exact classical method according to \equ{eq:d2Wclassical} and its non-relativistic approximation  \zit{eq:d2Wclassical_nonrel}, as well as the non-relativistic Schr\"odinger calculation\footnote{Note: The Schr\"odinger calculation only seems to exhibit far less noise than all the other calculations. The reason is that it was carried out using a much coarser frequency resolution (about one fifth). Apart from that, its prefactor is not known. To indicate this, the vertical scale in the corresponding plots was omitted.}, which is also based upon \equ{eq:d2Wqm_nonrel}, but is derived from a non-relativistic Schr\"odinger wave function.
As one can see, the three not entirely relativistic methods (``Dirac-Schr\"odinger'', ``classical theory (approx.)'', and ``Schr\"odinger'') significantly over-estimate the red shift\footnote{Note: These three methods even predict a red shift at $\vartheta=0$°, which is not shown here.}, because in these calculations, the Doppler effect due to the electron's drift in propagation direction is not included. The even harmonics, which should be visible for any non-zero observation angle, are missing in both spectra according to \equ{eq:d2Wqm_nonrel} (``Dirac-Schr\"odinger'' and ``Schr\"odinger''). This is because only the perpendicular component of the electric field is taken into account, weighted either with the Dirac or the Schr\"odinger probability density. As a result, the spectrum features a simple  $\cos(\vartheta)$-dependence and even vanishes for  $\vartheta \rightarrow 90$°. There is no change in structure with varying angle, it has always the shape of the version for $\vartheta = 0$°, although it is only valid for this particular angle.}

\english{Apart from the methodical differences laid out above, we also see differences between the two classical calculations and all three quantum mechanical ones. Especially the higher order harmonic peaks that are visible in the classical curves are averaged out in the case of the quantum spectra.}
\newpage
\section{Conclusions}
\label{sec:conclusions}
\english{From the above, it can be concluded that the procedure laid out in detail in subsections \ref{sec:qm_spectrum} to \ref{sec:qm_spectrum_limitations} is superior to the established techniques in terms of angle dependence of especially the even harmonic orders, and in terms of the predicted red shifts, whenever the parameters are chosen in a way that leads to a significant influence of retardation and magnetic field effects. In principal, even spin effects\footnote{We have chosen an initial state that has spin up with respect to the magnetic field direction $\sig{3}$, but did not analyze the influence of variations of the initial spin polarization on the spectrum. Spin effects such as those studied in \cite{Walser:1999} would even require three-dimensional calculations.} could be analyzed this way, since the method is based on the Dirac theory.
Although originally developed for the usage in combination with Dirac wave functions, the method could be employed in the context of the Schr\"odinger or Pauli theory as well\footnote{Note that in the Pauli theory, the charge current would have to be corrected for, see \autoref{sec:qm_spectrum_nonrel}.}, if one abandoned the Ehrenfest theorem and evaluated \equ{eq:ZE_Impulsintegral} directly, and it would at least correct the angle dependence. Employing \equ{eq:qm_spec} would even lead to correct red shifts, and the only remaining quantitative deviations would be due to the underlying non-relativistic wave packet propagation.
The computational cost for the spectrum alone is not essentially higher than for the more simple schemes, therefore its general usage can be recommended. However, the parameter regime in which its advantages are best visible are already very demanding as far as the wave packet propagation is concerned.}

\english{For the \emph {free} electron scenario, classical results are very similar to the ones obtained with our new method. We see no true quantum effects apart from the disappearance of the higher order harmonics. This may be attributed to averaging effects by the spatially extended Dirac charge current.
More significant effects are very likely to be observed once that \emph {atomic} systems are considered. It must be noted, though, that our main goal was to devise a relativistically correct scheme, as opposed to the non-relativistic existing ones. Our results show that the relativistic corrections are significant for the parameters employed here.}

\begin{ack}
The authors  wish to thank Andreas Staudt for his contribution of the Schr\"o\-din\-ger analysis.
\end{ack}
\appendix
\section{Classical equations of motion}
\label{chap:rk4}
\english{In classical relativistic mechanics, momentum change equals the applied force,} 
\begin{equation}
\vec  f= \frac{d \vec p}{dt}=\frac{d}{dt}\(\frac{m \vec v}{\sqrt{1-\frac{\vec v^2}{c^2}}}\)  \label{eq:BwGl}
\end{equation}
\english{where the force $\vec f$ is the well-known Lorentz-force:}
\begin{equation}
\vec f_{(\vec x, \vec v,t)} = q\vec E_{(\vec x,t)}+  \frac{q}{c}\vec v \times\vec B_{(\vec x,t)}.\label{eq:dpdt}
\end{equation}
\english{Equation \zit{eq:BwGl}, which represents a second-order differential equation with respect to the three-dimensional position variable $\vec x$ can be easily transformed \cite{KeitelSzymanowski:1998}} into a first-order system of differential equations in terms of the six-dimensional combined variable $(\vec x, \vec v)$:
\begin{eqnarray}
\frac{d}{dt}{\vec v}  &=& \frac{1}{m}\sqrt{1-\frac{\vec v\,^2}{c^2}} \(\vec f_{(\vec x, \vec v,t)} - \frac{1}{c^2} \(\vec v \cdot \vec f_{(\vec x, \vec v,t)} \)\vec v \)\label{eq:DGLsystem1}\\
\frac{d}{dt}{\vec x} &=& \vec v \label{eq:DGLsystem2}
\end{eqnarray}
\english{Formally, the above means}
\begin{equation}
\frac{d}{dt}{\(\vec x,\vec v\)}  = G_{(\vec x, \vec v,t)}\label{eq:DGLsytem}
\end{equation}
\english{where $G_{(\vec x, \vec v,t)}$ represents the six-dimensional function that is obtained from the right hand side  of equations \zit{eq:DGLsystem1} and \zit{eq:DGLsystem2} after inserting the force \zit{eq:dpdt}. The first-order system \zit{eq:DGLsytem} is then easily numerically integrated by means of a standard fourth-order Runge-Kutta algorithm \cite{Num_Rec}.}

\section{The Dirac equation}
\label{sec:Dirac_Hestenes}
\english{The Dirac equation in $3+1$ dimensions reads as follows:}
\onlypaper{%
\begin{equation}
\label{eq:Dirac_Hestenes}
i\hbar \ableit{\psi}{t} =  c\vec\alpha\cdot \(\frac{\hbar}{i}\vec\nabla-\frac{q}{c}\vec A\) \psi + \beta mc^2\psi +q A_0 \psi. 
\end{equation}}
\english{In the above, $\hbar$ is Planck's constant, $q$ and $m$ represent the particle's charge and rest mass, $c$ is the speed of light, $\alpha^i$ and $\beta$ are the usual Dirac matrices ($i=1,2,3$) \cite{Bjorken_Drell}. We use atomic units, so for an electron $\hbar=m=-q=1$~a.u., and $c=137.036$~a.u.}
\english{Suppose we have a potential that is invariant under translations in one particular direction that is designated by $\vec n$}%
\begin{equation}
  A_0=A_0(\vec x_\perp), \quad \vec  A=\vec A(\vec x_\perp),  \quad \text{\english{where }\deutsch{wobei }}\vec x_\perp=\vec n \times (\vec x \times \vec n),
\end{equation}
\english{then we can use the separation ansatz}
\begin{equation}
\psi (\vec x) = \psi'  (\vec x_\perp) \chi  (  x_\parallel)\quad \text{and}\quad
 \chi  (  x_\parallel)= \frac{1}{\sqrt{2\pi}} \exp\( \frac{i}{\hbar}p_\parallel^0 x_\parallel\), \quad \text{\english{where }\deutsch{wobei }} x_\parallel = \vec n \cdot \vec x \label{eq:2dAnsatz}
\end{equation}
\english{in the Dirac equation \zit{eq:Dirac_Hestenes}. The result is:}%
\begin{equation}
i\hbar \ableit{\psi'}{t} =  c\vec\alpha\cdot \(\frac{\hbar}{i}\vec\nabla_\perp + p_\parallel^0\vec n- \frac{q}{c}\vec A\) \psi' + \beta mc^2\psi' +q A_0 \psi', \, \text{\english{where }\deutsch{wobei }} \vec\nabla_\perp=\vec n \times (\vec \nabla \times \vec n).
\end{equation}
\english{In general, there will not be a single momentum $p_\parallel^0$, but rather a momentum distribution giving specific weights for arbitrary $p_\parallel \in \real$ and the solution is a wave packet instead of \zit{eq:2dAnsatz}. We restrict ourselves to the simple choice ${p_\parallel^0=0}$ to eliminate this constant. Other values would not provide more physical insight, and true wave packet calculations would cause the computing times to grow into the regime of true three-dimensional calculations. Note however, that the ansatz \zit{eq:2dAnsatz} inevitably leads to a meaningless factor $\hbar \delta(0)$ in bilinear expressions of $\psi$, such as expectation values or the integrated current density. 

We finally obtain an effective Dirac equation in two dimensions:}%
\begin{equation}
i\hbar \ableit{\psi'}{t} =  c\vec\alpha\cdot \(\frac{\hbar}{i}\vec\nabla_\perp - \frac{q}{c}\vec A\) \psi' + \beta mc^2\psi' +q A_0 \psi'  \label{eq:Dirac2D}
\end{equation}
\english{The above equation is integrated \cite{Rathe:1997} using the split-operator scheme \cite{Braun:1999} presented in our earlier work \cite{Mocken:2003b}. The dynamical grid techniques described in there are especially well-suited for the free electron problem discussed here, and dramatically reduce computation times.}

\english{A vector potential that describes a laser beam of electric field amplitude $E_0$ and frequency $\omega$,  which is linearly polarized in direction $\hat a$ and propagates in direction of the wave vector $\vec k$, such as}%
\begin{eqnarray}
\vec A = -\hat a \frac{c}{\omega} E_0\, \sin \,(\omega t - \vec k \cdot \vec x), \quad A_0 =0, \label{eq:laserpotential}
\end{eqnarray}%
\english{is invariant in direction $\vec n \ovdefeq \hat k \times \hat a$. As can be easily seen, in this context ``two-dimensional'' does not mean that there is no magnetic field $\vec B$ (which points in the $x_\parallel$-direction). Also note that the above two-dimensional treatment is \emph{exact} for the special case of a free electron in a laser field that we considered in this paper, because the vector potential \zit{eq:laserpotential} features the required symmetry.}%

%
%
\bibliographystyle{elsart-num} 
\bibliography{../database}
\end{document}